\documentclass{aa}
\usepackage{nccmath}
\usepackage{graphicx}
\usepackage{txfonts}
\usepackage{natbib}
\bibpunct{(}{)}{;}{a}{}{,}

\begin{document}

\title{Role of the impact parameter in exoplanet transmission spectroscopy}
\author{X.~Alexoudi\inst{1,2,3}
          \and
          M.~Mallonn \inst{1}    
        \and
          E. Keles \inst{1,3}
          \and
          K. Poppenh{\"a}ger \inst{1,3}
          \and
            C.~von Essen \inst{4}
           \and
         K.\,G.~Strassmeier \inst{1,3}  \\           
          }

   \institute{Leibniz-Institut f\"{u}r Astrophysik Potsdam (AIP), An der Sternwarte 16, D-14482 Potsdam, Germany \\
\email{xalexoudi@aip.de}
         \and 
         Potsdam Graduate School, Am Neuen Palais 10, 14469 Potsdam, Germany
         \and 
         University of Potsdam, Am Neuen Palais 10, 14469 Potsdam, Germany
         \and 
        Stellar Astrophysics Centre (SAC), Department of Physics and Astronomy, Aarhus University, Ny Munkegade 120, DK-8000, Aarhus C, Denmark
        }

   \date{Sept 23 2019}


\abstract{Transmission spectroscopy is a promising tool for the atmospheric characterization of transiting exoplanets. Because the planetary signal is faint, discrepancies have been reported regarding individual targets. 
  
}{
We investigate the dependence of the estimated transmission spectrum on deviations of the orbital parameters of the star-planet system that are due to the limb-darkening effects of the host star. We describe how the uncertainty on the orbital parameters translates into an uncertainty on the planetary spectral slope.
}{We created synthetic transit light curves in seven different wavelength bands, from the near-ultraviolet to the near-infrared, and fit them with transit models parameterized by fixed deviating values of the impact parameter $b$. First, we performed a qualitative study to illustrate the effect by presenting the changes in the transmission spectrum slope with different deviations of $b$. Then, we quantified these variations by creating an error envelope (for centrally transiting, off-center, and grazing systems) based on a derived typical uncertainty on $b$ from the literature. Finally, we compared the variations in the transmission spectra for different spectral types of host stars. 
}{Our simulations show a wavelength-dependent offset that is more pronounced at the blue wavelengths where the limb-darkening effect is stronger. This offset introduces a slope in the planetary transmission spectrum that becomes steeper with increasing $b$ values. Variations of $b$ by positive or negative values within its uncertainty interval introduce positive or negative slopes, thus the formation of an error envelope. The amplitude from blue optical to near-infrared wavelength 
for a typical uncertainty on $b$ corresponds to one atmospheric pressure scale height and more.
This impact parameter degeneracy is confirmed for different host types; K stars present prominently steeper slopes, while M stars indicate features at the blue wavelengths.    
}{We demonstrate that transmission spectra can be hard to interpret, basically because of the limitations in defining a precise impact parameter value for a transiting exoplanet. This consequently limits a characterization of its atmosphere. 
}

\maketitle
\section{Introduction}

Studying transiting exoplanets has been one of the highlights of the past 20 years in astronomy. Since the development of suitable instrumentation and techniques, the transit events permit us to even probe the atmospheres of exoplanets and allow a glimpse in their interiors \citep[e.g.,][]{Charbonneau2002}. One approach for exploring the atmosphere is employing of low-resolution transmission spectroscopy, which is a very effective method for investigating large gas giant exoplanets, such as the so called hot Jupiters. Its principle is that the effective radius of a planet depends on the wavelength. This wavelength can be measured during a planet transit (planet passes in front of its host star), and the planetary atmosphere can be seen to interact with the starlight. When we measure the ratio of the planet-to-star radius over wavelength, we obtain a transmission spectrum \citep[e.g.,][]{Kreidberg18}. Transmission spectra can reveal a wealth of features in an atmosphere, such as signatures of Rayleigh scattering toward shorter wavelengths, which is attributed to aerosols or $H_2$, and clouds, atomic, and molecular absorption from Na and K, H$_{2}$O, AlO, TiO, or VO \citep[e.g.,][]{sing2016,MallonnStrass,tsiaras2018,kreidberg2018,nikolov2018, mancini2019, vonessen2019}. One drawback is that the spectral slope can be mimicked by other effects, for instance, third-light contamination from a stellar companion or the  potential activity of the host star, for example, spots and faculae \citep{McCullough2014,oshagh2014, rackham2018}. It is difficult to interpret the slope at optical wavelengths, and it can result in inconsistencies in the atmospheric characterization of exoplanets. 

The literature lists several cases of independently derived transmission spectra of individual targets that deviate significantly. Numerous authors suggested that these reported discrepancies might be solved by a homogeneous reanalysis of the individual datasets to avoid systematic effects originating from differences in data reduction or data analysis. A successful application was presented by \cite{Alexoudi2018}, who reanalyzed the data of two independent studies on the hot Jupiter HAT-P-12b. The two previous investigations, \cite{mallonn2015} and \cite{sing2016}, derived inconsistent conclusions on the planetary atmosphere; the former found a flat optical spectrum, while the latter concluded that HAT-P-12b has a Rayleigh scattering slope toward the blue wavelengths. The probable source of this deviation was the use of different values of the orbital inclination in the two analyses. The difference of these inclination values was about $2.2\sigma$. By applying a common inclination value in a simultaneous transit fit to the acquired datasets (both ground- and space-based), all data yielded consistent results. The authors finally concluded that weak scattering at short optical wavelengths is present in the planetary atmosphere. 

Motivated by the discovery of this effect in HAT-P-12b transmission spectra, we present in this work an extended investigation of this phenomenon. With noise-free, simulated light curves, we prove here that the limited precision in the knowledge of the transit parameters can affect the transmission spectra, and lead into misinterpretations and discrepancies in the literature. 

In Section 2 we present the methods that were employed in our work. In Section 3 we address the effect of the orbital parameters on the transmission spectra with simulations and through two different approaches. This confirms a degeneracy with the spectral slope. Then, we investigate the extension of this effect on a sample of hot Jupiters with different types of host stars. In Section 4 we aim to explain some of the known discrepancies from the literature that are due to the aforementioned effect, and in Section 5 we summarize our work and present our final conclusions.

\section{Methods}

One of the fundamental steps in a light-curve analysis is fitting a transit model on a given dataset. Typically, the transit-model fit parameters are the orbital period of the system $P$, the inclination $i$, the semimajor axis in units of stellar radii $a/R_s$, the transit depth in terms of the ratio of planet to stellar radii ($R_p/R_s$), the limb-darkening (LD) effect, expressed using different LD laws (LDL) and coefficients (LDCs), the midtime of the transit and the contribution of a third light term in the light curve. 

Investigations of low-resolution transmission spectroscopy often start with the analysis of a white-light curve, which is the light integrated over the entire observed wavelength range. Such a light curve normally holds a low value of photon noise and therefore presents high photometric precision. This precision allows determining the entire set of model parameters. In the next step, the observed wavelength range is split into numerous wavelength channels to create a chromatic set of light curves and investigate the wavelength dependence of transit parameter of interest, here mainly $R_p/R_s$. Because the chromatic light curves are of lower photometric precision, all parameters that are not expected to vary with wavelength are normally kept fixed to their values derived in the previous white-light curve fit. These achromatic parameters also include those describing the planetary orbit. The planetary trajectories around their host stars can be tracked with the use of the impact parameter $b$. This is the relation between the orbital inclination and the semimajor axis, according to the following expression \citep{Winn2009,Haswell2010}: 

\begin{ceqn}
\begin{align}
    {{b} = \frac{cos(i) \times a}{R_s.}\, }
    \label{impact_parameter}
\end{align}
\end{ceqn}
However, the observed uncertainties in $i$ and $a/R_s$ allow for a range of $b$ values. If the uncertainty in $b$ is large, then the planet might apparently follow different pathways over its host, and because the stellar surface does not have a homogeneous brightness, the effect of the LD becomes important. 
We are interested in studying whether this range in allowed pathways over the host star might result in an uncertainty of the derived values of $R_p/R_s$ over wavelength, that is, the planetary transmission spectrum. Eventually, we wish to describe the extension of this effect with respect to the atmospheric pressure scale height $H$ of an exoplanetary atmosphere. The scale height is a quantity according to which we can estimate the size of the absorbing annulus of the planetary atmosphere, as defined by the expression \citep{Winn2010}

\begin{ceqn}
\begin{align}
{{H} = \frac{k_B \ T_{eq}}{\mu_m \ g },}
\label{scale_height}
\end{align}
\end{ceqn}
where $k_B$ is the Boltzmann constant, $T_{eq}$ is the equilibrium temperature, $\mu_m$ is the mean molecular mass, and g is the local gravitational acceleration. 

\section{Results }
\subsection{Impact parameter degeneracy}
\label{sec_res1}

Ideally, individual investigations of the atmosphere of the same exoplanet would yield consistent results, but this is not always the case: the literature includes reported discrepancies regarding the atmospheric characterization of exoplanets. When the authors do not agree at the same employed parameter values in the light-curve analysis, then inconsistent transmission spectra might appear. This is especially evident when different works use very different orbital parameters of $i$ and $a/R_s$, and hence a different $b$ value. In this section we focus on the effect of the choice in $b$ on synthetically retrieved transmission spectra of transiting exoplanet events, and more specifically, their spectral slopes. We address the problem of the effect on the spectral slope when the analysis involves fixed orbital parameters, $i$ and $a/R_s$, on values that yield different impact parameters for the system.

For this purpose, we simulated noise-free light curves in multiple wavelength bands by assuming a hot Jupiter exoplanet that is on a circular orbit of a K-type host star with T\textsubscript{eff}=4500K, surface gravity of log $\textit{g}$ = 4.5, and solar metallicity. The orbital period is set to 3.32 d. The orbital inclination of the system is $i=90 ^{\circ}$, the semimajor axis is $a$ = 8 $R_s$, and the planet transits its host centrally at a trajectory defined by an impact parameter $b$ = 0. We used the four-parameter LD law to simulate the stellar LD, with coefficients from \cite{Claret2011}. We defined a transit depth of $2\%$ ($R_p/R_s = 0.14142$). 
The synthetic light curves were created with a custom pipeline using PyAstronomy\footnote{https://github.com/sczesla/PyAstronomy} and the analytical transit models of \cite{mandel&agol2002}. We worked with sets of seven chromatic light curves of different bands (Johnson/Cousin U, B, V, R, I, J, and H) and we did not consider any out-of-transit variations. 

We focused on the consequences that the forced alterations of the orbital parameter values might have on the derived transmission spectra when their combination yields different or similar $b$ values. We therefore kept $i$ and $a/R_s$ fixed to values that deviate from their original ones during the application of a transit model fit. We used the OneDFit class, which is an object-fitting base class of PyAstronomy. It provides a suitable interface for the Nelder-Mead simplex, which is a parameter-fitting algorithm, in order to determine the best-fit solution. In Table \ref{tab_Fig1} we present the values we adopted to create the light curve and the altered values that were kept fixed at the subsequent model fitting. The only free parameters during the fit were $R_p/R_s$ and the orbital period $P$. To illustrate the individual effects of deviations in $i$, $a/R_s$, and $b$ on the retrieved transmission spectrum, we varied $i$ and $a/R_s$ by unusually high values, much higher than their typical uncertainties. It was therefore necessary to also vary $P$ accordingly to achieve a reasonable model fit to the simulated data. This was done by including $P$ in this part of the work as free-to-fit parameter.

\begin{figure}[ht]
\centering
\includegraphics[height=6.1cm]{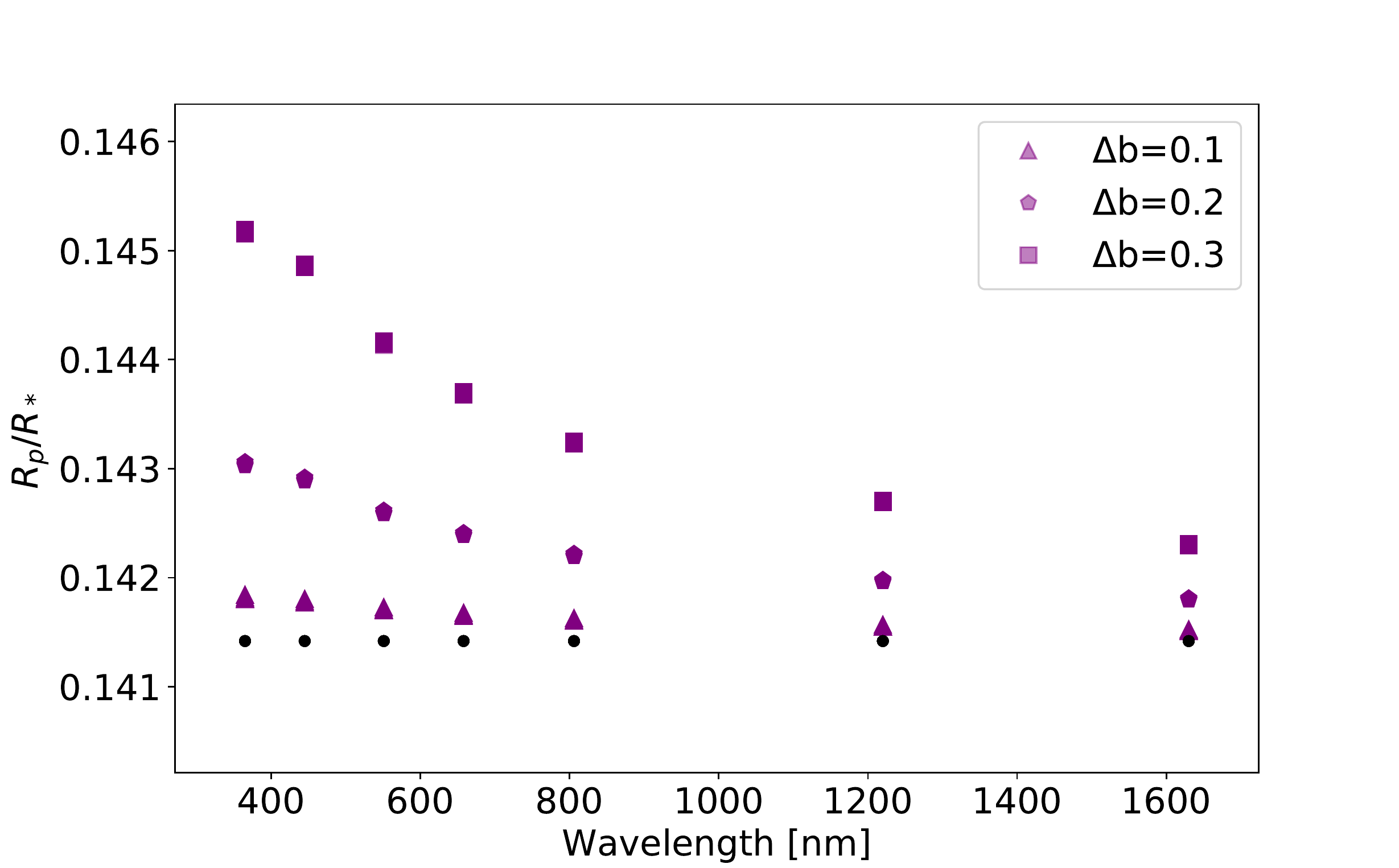}
\caption{Synthetic transmission spectra for transiting exoplanets. We show the spectral slopes derived with fixed orbital parameters in combinations that yield the same impact parameter. The symbols indicate the different variations in $b$. The nine configurations of Table\ref{tab_Fig1} overlap in three sequences according to their $\Delta b$ deviation from the original setup (black dots).}
\label{synthetic_transmission_spectra_qualitative}
\end{figure}

In Fig.\ref{synthetic_transmission_spectra_qualitative} we present the transmission spectra derived with this approach. They clearly show a wavelength-dependent offset in $R_p/R_s$. The nine parameter sets of Table \ref{tab_Fig1} overlap in three sequences, indicating the parameter sets with the same $b$ value. They are distinguished by different symbols for each $\Delta b$ configuration. Even when the values for $i$ and $a/R_s$ are very different, we obtain the same offset if the parameter combinations result in the same $b.$ This can be compared to the initial setup value of the simulations for $R_p/R_s$. With this qualitative approach, we wish to highlight that this wavelength-dependent offset in $R_p/R_s$ depends mainly on $b$, not on $i$ or $a/R_s$ separately, and that it is more pronounced toward shorter wavelengths. This introduces a slope in the spectra. This offset has a strong nonlinear exponential dependence with the deviation in $b,$ and it is basically driven by the LD of the host star. 

The offset in the derived $R_p/R_s$ can be explained as follows: When we fix $b$ to a different value than the true input value, the planet is forced toward a deviating trajectory. The fit compensates for the different brightness of the host star that is due to LD along this trajectory by a planetary radius $R_p$ different from the input radius, causing the offset. In synthetic light curves, where the host star LD is switched off, no offset in $R_p/R_s$ is found. Hereafter, we call this effect the "impact parameter degeneracy". 

\begin{table}
\caption{Setup of the orbital parameters $i$ and $a$/$R_{star}$ of the transit light-curve simulations presented in Fig. \ref{synthetic_transmission_spectra_qualitative}}
\label{tab_Fig1}
\begin{center}
\begin{tabular}{cccc}
\hline
\hline
\noalign{\smallskip}
     &  $i$ & $a/R_{star}$ & $b$  \\
\hline
\noalign{\smallskip}
simulated &  90  &   8  & 0      \\
\hline
fitted   &  89.28 & 8  & 0.1    \\
         &  88.57 & 8  & 0.2    \\
         &  87.85 & 8  & 0.3    \\
         &  89.05 & 6  & 0.1    \\
         &  88.09 & 6  & 0.2    \\
         &  87.13 & 6  & 0.3    \\
         &  89.43 & 10 & 0.1    \\
         &  88.85 & 10 & 0.2    \\
         &  88.28 & 10 & 0.3    \\
\hline 
\end{tabular}
\end{center}
\end{table}

\begin{figure}[ht]
\center
\includegraphics[height=6cm]{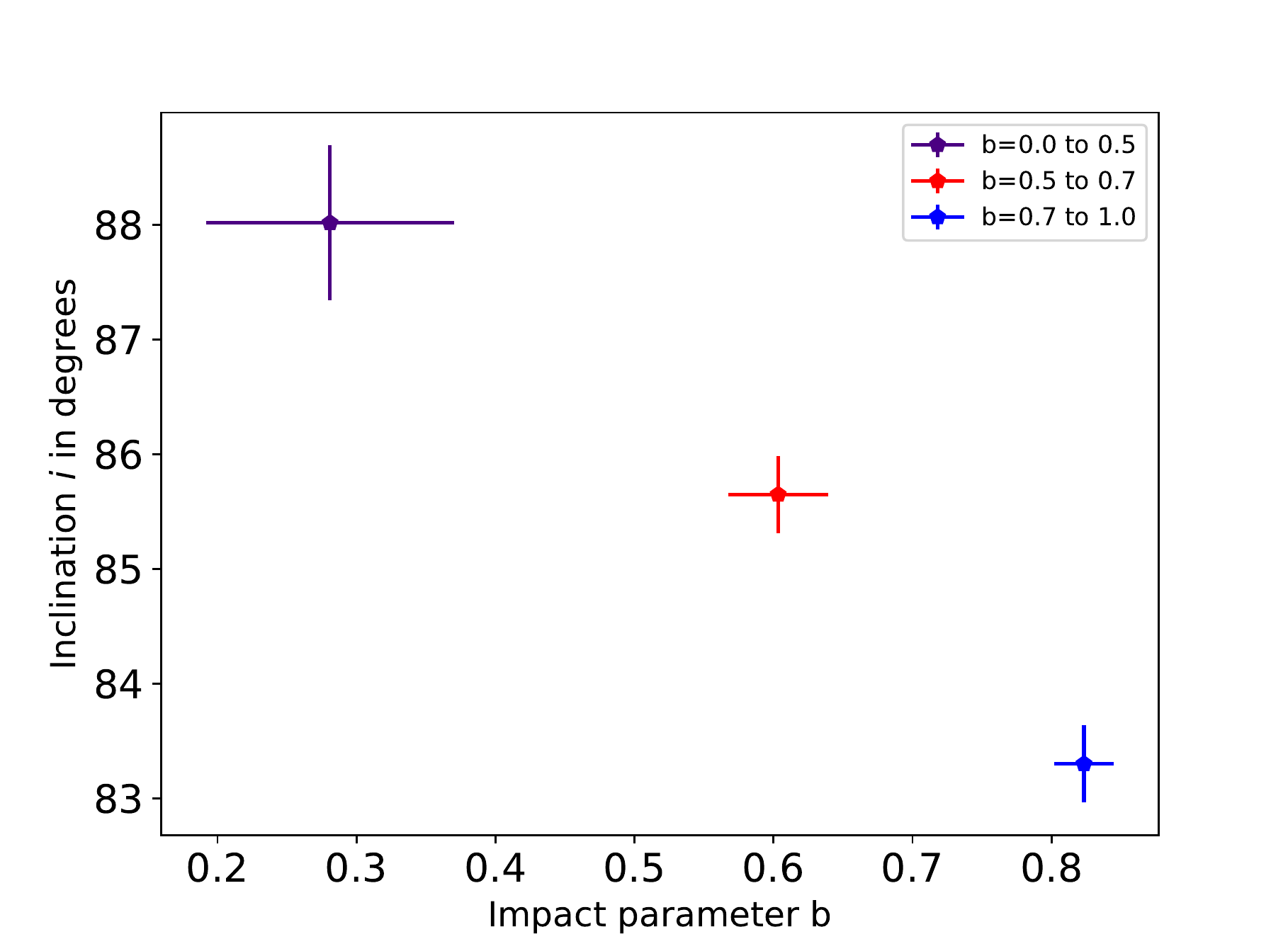}
\caption{Median inclination values and their uncertainties with respect to the median $b$ values and their uncertainties in $b$ of the different subgroups of exoplanets. Purple values represent group A, red values group B, and blue values group C}.
\label{new_sample_hot_jupiters_uncertainties}
\end{figure}

\begin{figure}[ht]
\includegraphics[height=6.5cm]{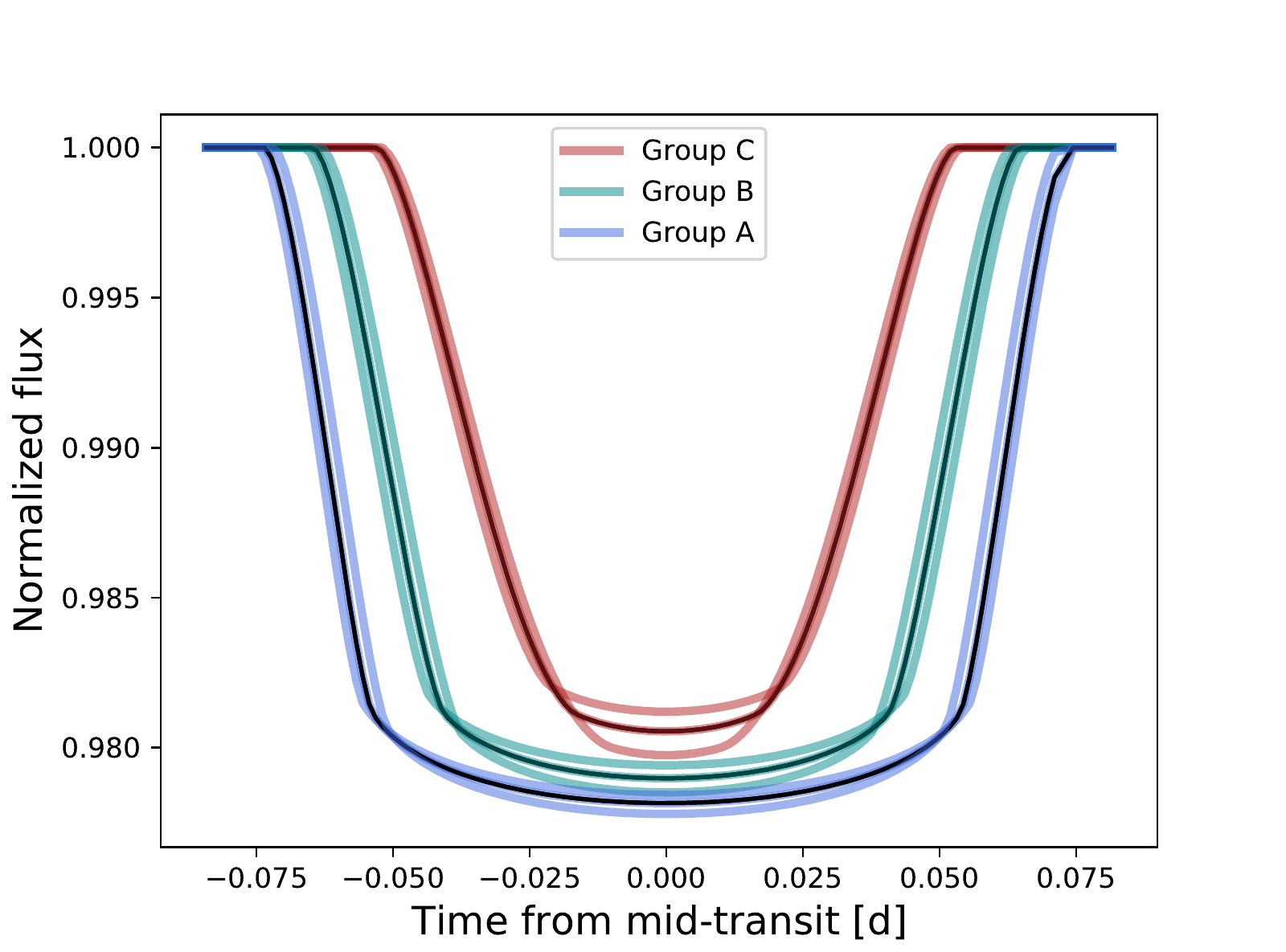}
\caption{Example of the synthetic light-curve fit for each subgroup.}
\label{lcs_fits}
\end{figure}

\begin{figure*}[ht]
\centering
\includegraphics[height=8.5cm]{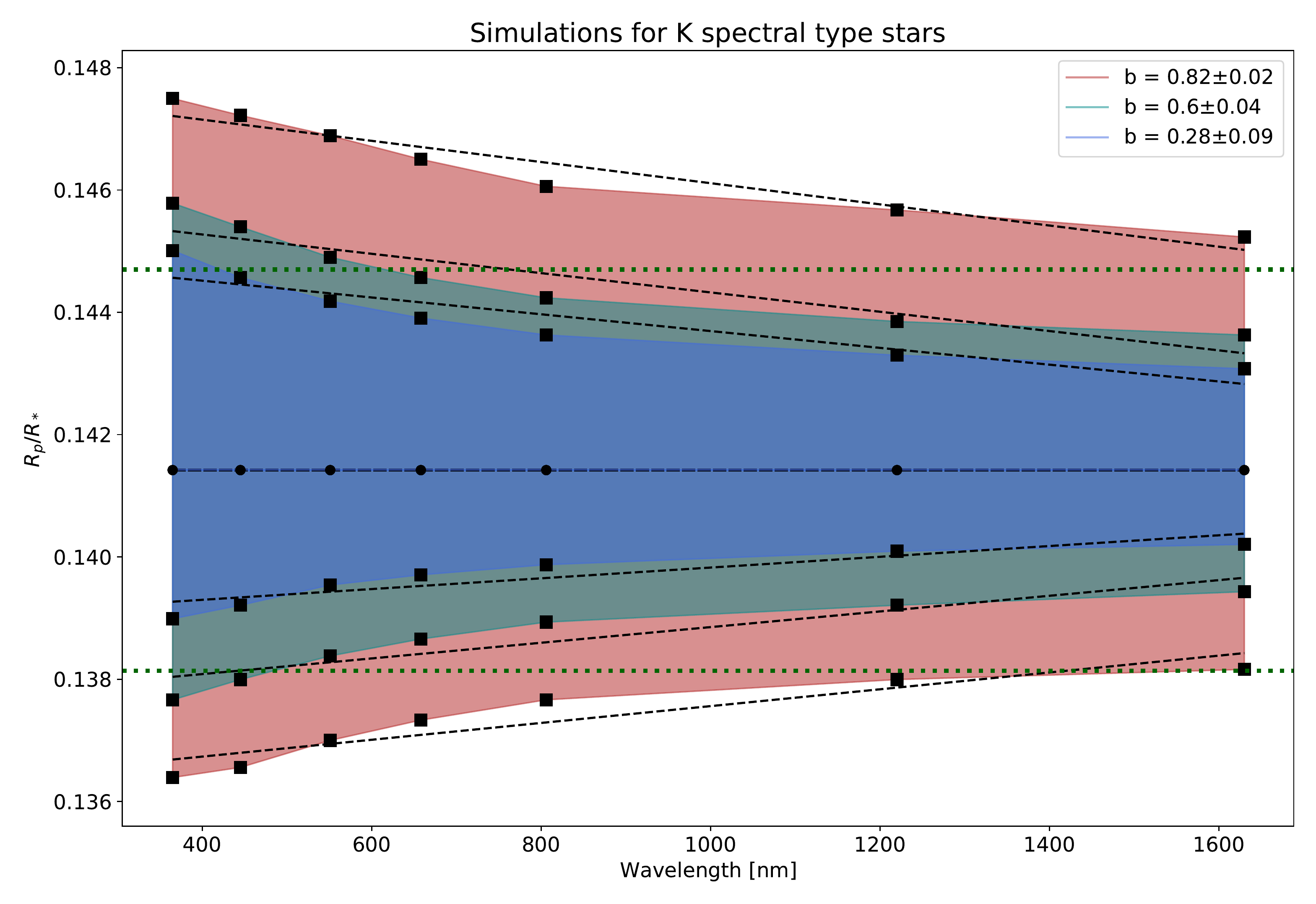}
\caption{Effect of $b\pm \Delta b$ on the transmission spectra of three different groups of exoplanets (groups A, B and C), showing an introduced slope and an offset for different $b$ values. Black dots show the synthetic spectra of each subgroup, and black squares show the respective derived spectra with the variation in $b$. The colored areas illustrate the error envelope for each case. Dashed black lines show the linear regression fits on each spectrum, and dotted green lines indicate two atmospheric scale heights from the predefined input value for $R_p/R_s$.}
\label{transmission_spectra_deltab}
\end{figure*}

\subsection{Formation of an uncertainty envelope}
\label{sec_res2}

The uncertainty in $b$ is able to modify the transmission spectrum, as we demonstrated in the previous paragraph. In order to quantify the effect of a typical uncertainty in $b$ on the transmission spectra through the impact parameter uncertainty, we obtained the impact parameter and its uncertainty for a total sample of 349 hot Jupiters from the NASA Exoplanet Archive \citep{Akeson2013}. We split our sample into three groups of interest according to their $b$ values: centrally to off-center (group A includes 181 exoplanets with $0.0 < b < 0.5$),  off-center to nearly grazing (group B includes 82 exoplanets with $0.5<b<0.7$), and totally grazing (group C includes 62 exoplanets with $0.7<b<1$) transiting exoplanets. For each group, we derived the median parameter values and median uncertainties and show them in Fig. \ref{new_sample_hot_jupiters_uncertainties}. 
The uncertainty $\Delta b$ decreases with increasing $b$ because $\Delta i$ is determined more precisely for higher $b$ values. The $b$ values and their median uncertainties that we used here as typical $\Delta b$ for each subgroup are $b_A=0.28\pm0.09$, $b_B=0.6\pm0.04,$ and  $b_C=0.82\pm0.02$. 

To simulate the effect of the typical impact parameter uncertainty, we created one synthetic light curve per group and per filter with the median $b$ and performed a transit model fit by fixing $b$ to values corresponding to plus and minus the median $\Delta b$. To this end, we always fixed $a/R_s=8$ and adopted $i$ to yield the demanded $b$ value. The orbital period was again set to 3.32 days, and the 
LD was treated using the four-parameter law. The adopted LDCs from \cite{Claret2011} correspond to an average K-type host star from our sample with T\textsubscript{eff}=5000K, surface gravity of log $\textit{g}$ = 4.5, and solar metallicity. The only parameter that was free to fit per light curve was $R_p/R_s$. In contrast to our exercise in Section~\ref{sec_res1}, we varied $i$ by only low values and kept the parameter $P$ fixed to its input value. This reflects the typical situation of $P$ being fixed during the fit of the multiwavelength transit light curves because it is usually known to high precision. An example of the fits of these synthetic light curves is presented in Fig. \ref{lcs_fits}. 

The results show the deformations of the spectra that are caused by changes in $b$ of each subgroup; they are presented in Fig. \ref{transmission_spectra_deltab}.The deviation in $b$ in opposite directions, according to plus and minus the uncertainty, result in an opposite offset in $R_p/R_s$, and the formation of an uncertainty envelope. The deviation in $b$ towards a centrally or grazing transiting configuration (lower or higher $b$ value) results in a positive or negative slope in the transmission spectrum. We calculated the slope values, which are equal to $-1.37\,\pm\,0.27\,\times\,10^{-6}$~nm$^{-1}$ for group A,  $-1.58\,\pm\,0.31\,\times\,10^{-6}$~nm$^{-1}$ for group B, and  $-1.73\,\pm\,0.23\,\times\,10^{-6}$~nm$^{-1}$ for group C, for the effect of $+\Delta b$. Because the uncertainty envelope closes with the effects due to $-\Delta b$, the estimated slope values of the opposite direction are $0.88\,\pm\,0.19\,\times\,10^{-6}$~nm$^{-1}$ for group A,  $1.28\,\pm\,0.25\,\times\,10^{-6}$~nm$^{-1}$ for group B, and  $1.38\,\pm\,0.26\,\times\,10^{-6}$~nm$^{-1}$ for group C.
The slopes clearly increase slightly from a central toward a grazing transit geometry. 

In the physical interpretation of the derived transmission spectrum, a rise of $R_p/R_s$ toward shorter wavelengths can be conceived as scattering of a hazy atmosphere, probably due to small particles \citep[e.g.,][]{Pont2013,MallonnWakeford2017,MacDonald2020}. We conclude that a flat spectrum (initial assumption of our simulation) might appear sloped and can be misinterpreted as Rayleigh absorption. Moreover, the opposite, a flat spectrum, might be the outcome of using a lower $b,$ and a plausible Rayleigh feature is obscured with such a configuration. Alternatively, a planet without an atmosphere can be considered to have an atmosphere for the same reasons as previously because a spectral slope appears at shorter wavelengths, which is driven by the LD effect in the host star, and is due to a poor knowledge in $b$. 

In transmission spectroscopy, the variation in $R_p/R_s$ with wavelength is usually expressed in units of the atmospheric pressure scale height $H$. In order to examine the impact parameter degeneracy on the spectral slope in units of $H$, we determined a typical value of $H$ for hot Jupiters that is suitable for transmission spectroscopy. We ranked all planets from the TEPCat \citep{Southworth2011} according to the amplitude of their potential transmission signal $\Delta_\delta$, estimated by \cite{Winn2010} to

\begin{ceqn}
\begin{align}
{{\Delta_\delta} = \left(\frac{R_{p} + N_{H} \ H }{ R_{s}}\right)^2-\left(\frac{R_{p}}{ R_{s}}\right)^2,}
\label{transmission_signal}
\end{align}
\end{ceqn}
with $N_H$ as the number of scale heights, set to 1 for the purpose of our ranking. 
Then we formed a typical value by the average of the top-ranked 30 objects. The atmosphere of a typical hot Jupiter suitable for transmission spectroscopy causes a $\Delta_\delta$ signal of $\sim 4.5 \times 10^{-4}$ , and planets with smaller atmospheric scale heights cause significantly weaker signals. We obtained an average stellar radius of our sample of $R_s$=1$R_\odot$ and an average value for the atmospheric scale height of $H$=1140 Km. We conclude that a representative relative scale height of our sample is $h$=$H/R_s$=0.00164. In Fig. \ref{transmission_spectra_deltab}, we show two atmospheric scale heights of the average $R_p/R_s$ input value (green dotted lines).  
Intriguingly, the relative $R_p/R_s$ change over wavelength of the spectral slope is about one atmospheric scale height for hot Jupiters with a strong signal $\Delta_\delta$. The same amplitude of the impact parameter degeneracy corresponds to even higher values in units of $H$ for exoplanets with weaker transmission signals.   

\begin{figure*}[ht]
\includegraphics[height=7.5cm]{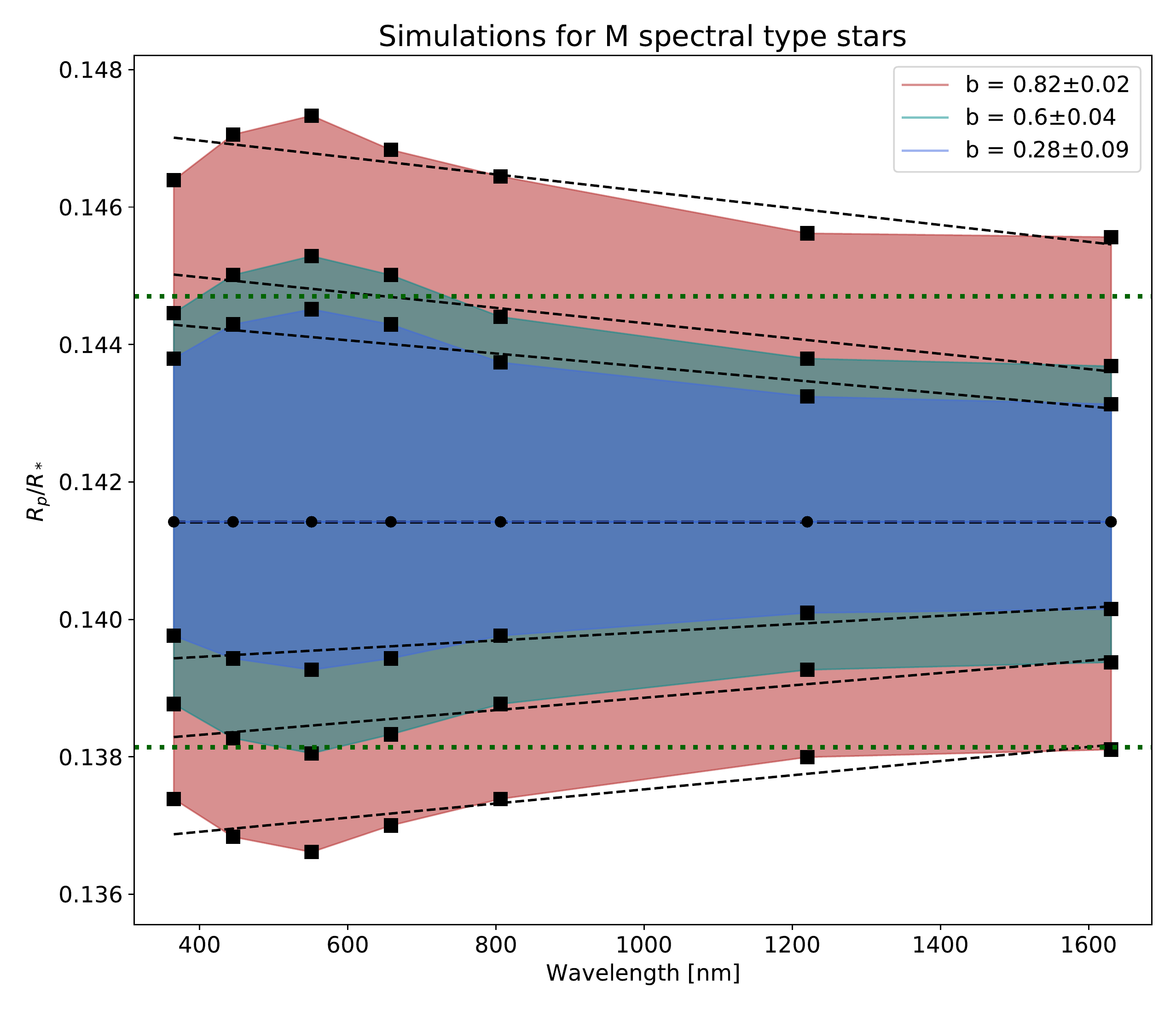}
\includegraphics[height=7.5cm]{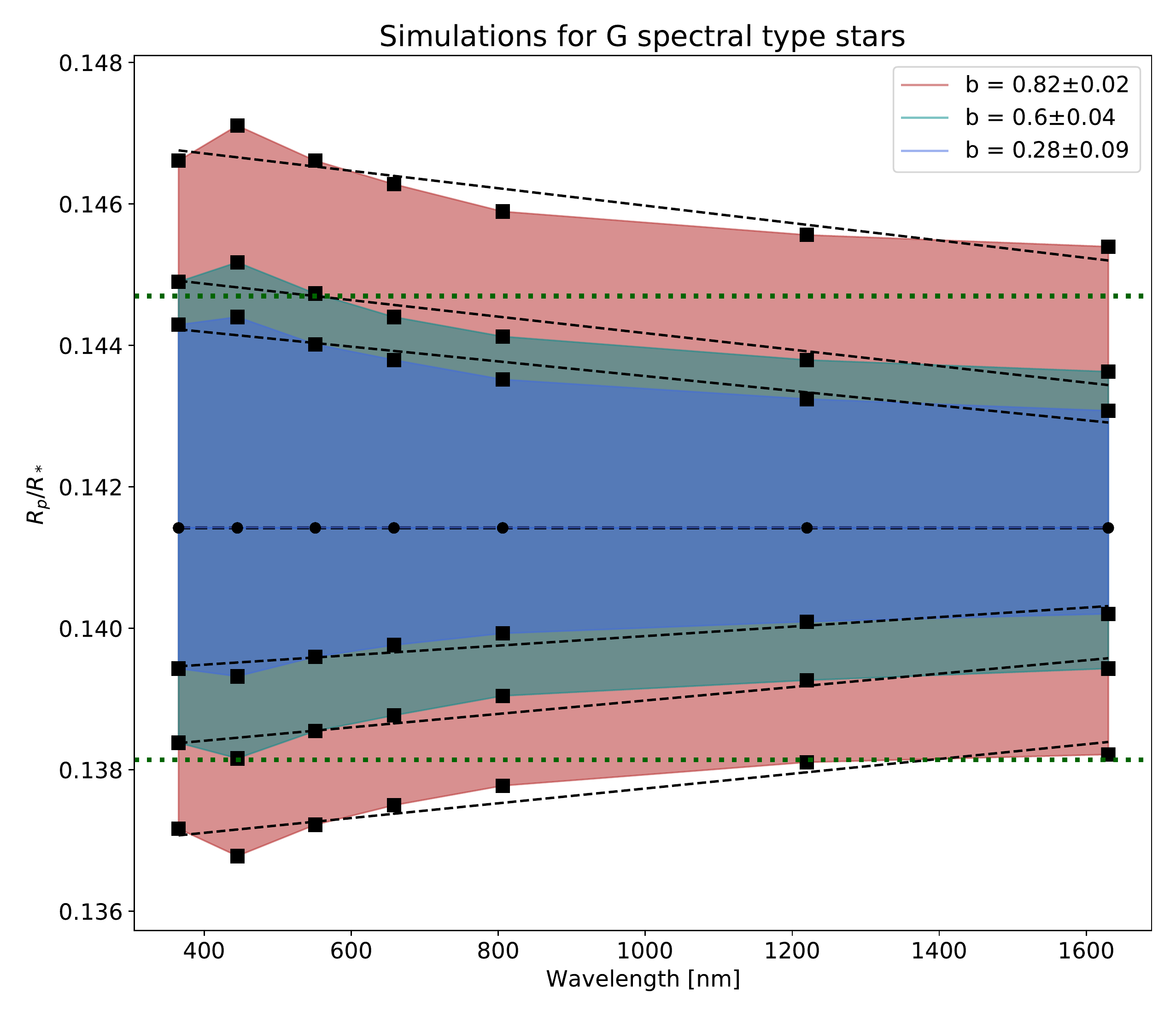}
\includegraphics[height=7.5cm]{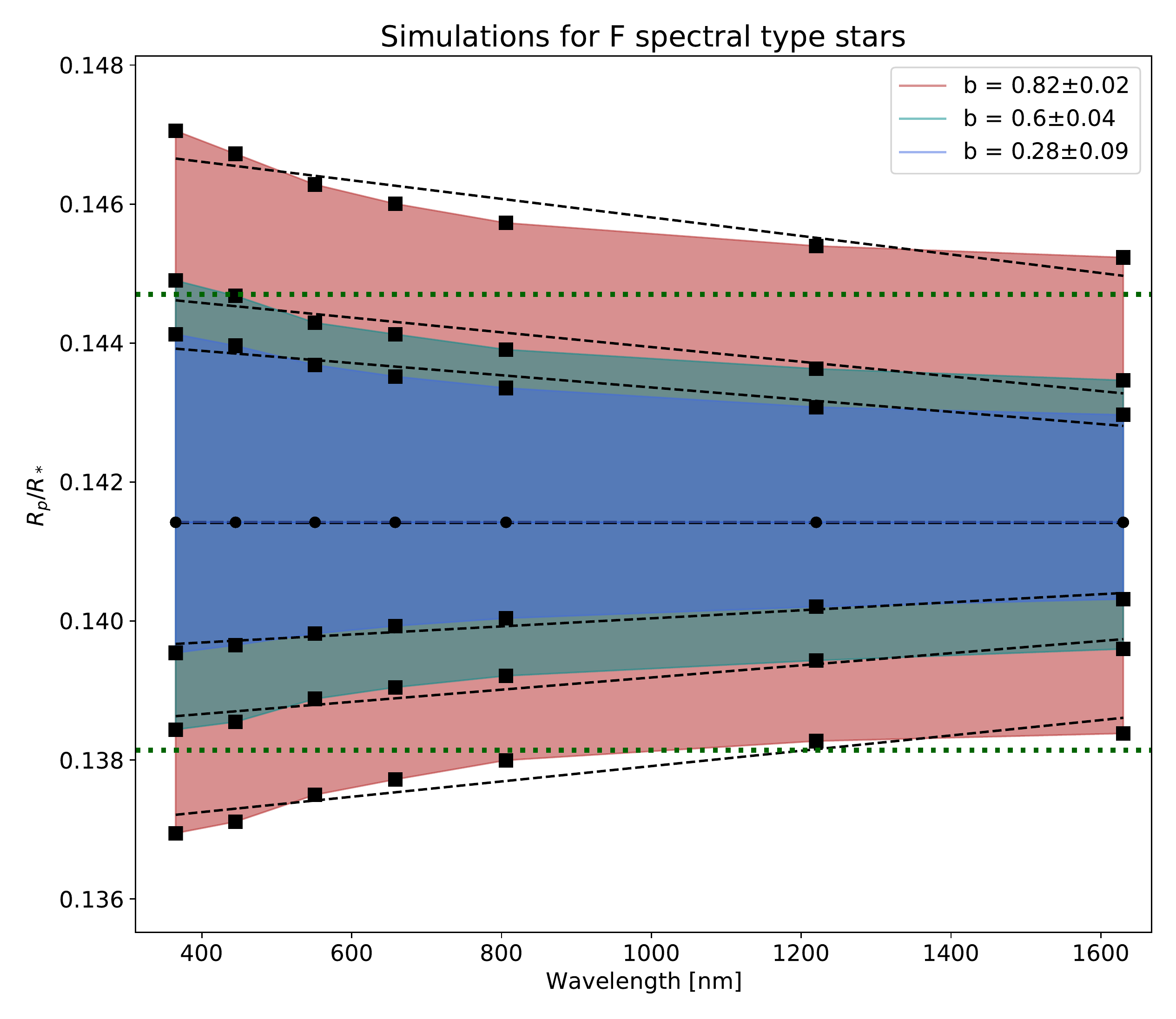}
\includegraphics[height=7.5cm]{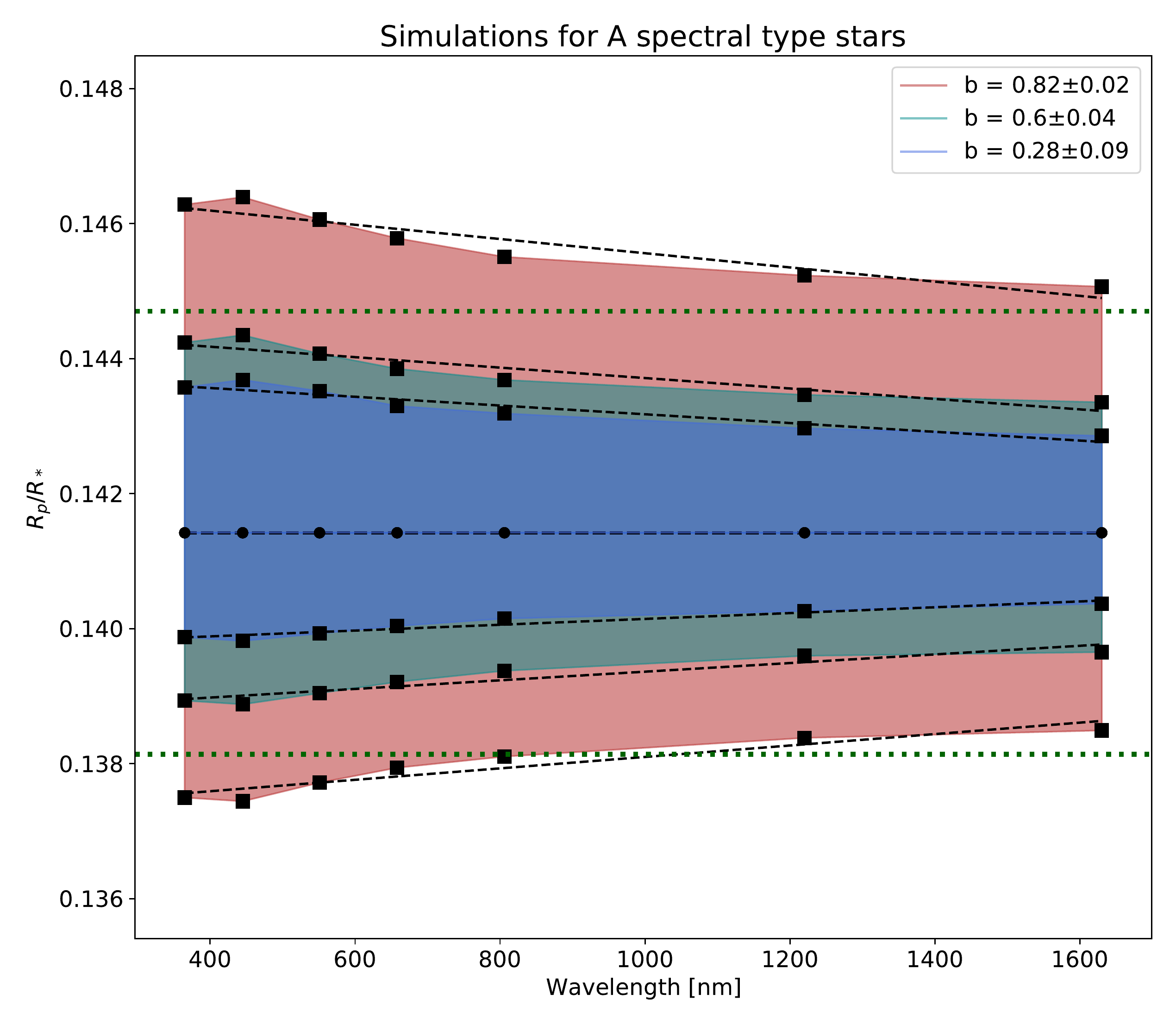}
\caption{Simulations of variations that are due to the impact parameter degeneracy in the transmission spectra of the three different subgroups of exoplanets (groups A, B, and C), orbiting different categories of host stars. The upper panels show M- (on the left) and G- (on the right) and the lower panels F- (on the left) and A-type stars (on the right). The black dots indicate the flat synthetic spectra of each subgroup that we created with those configurations of the orbital parameters that yielded the median $b$ value for each case. A combination of the orbital parameters at the transit model fit yields the spectra based on the change of $\pm \Delta b$ for each subgroup; this is indicated with black squares. The colored areas represent the error envelope. Dashed black lines show the linear regression fits on each spectrum, and the dotted green lines indicate two atmospheric scale heights from the average $R_p/R_s$ value, as defined from the original setup for the transit depth.}
\label{diff_spectral_type_hosts}
\end{figure*}

\begin{figure}[ht]
\includegraphics[height=7cm]{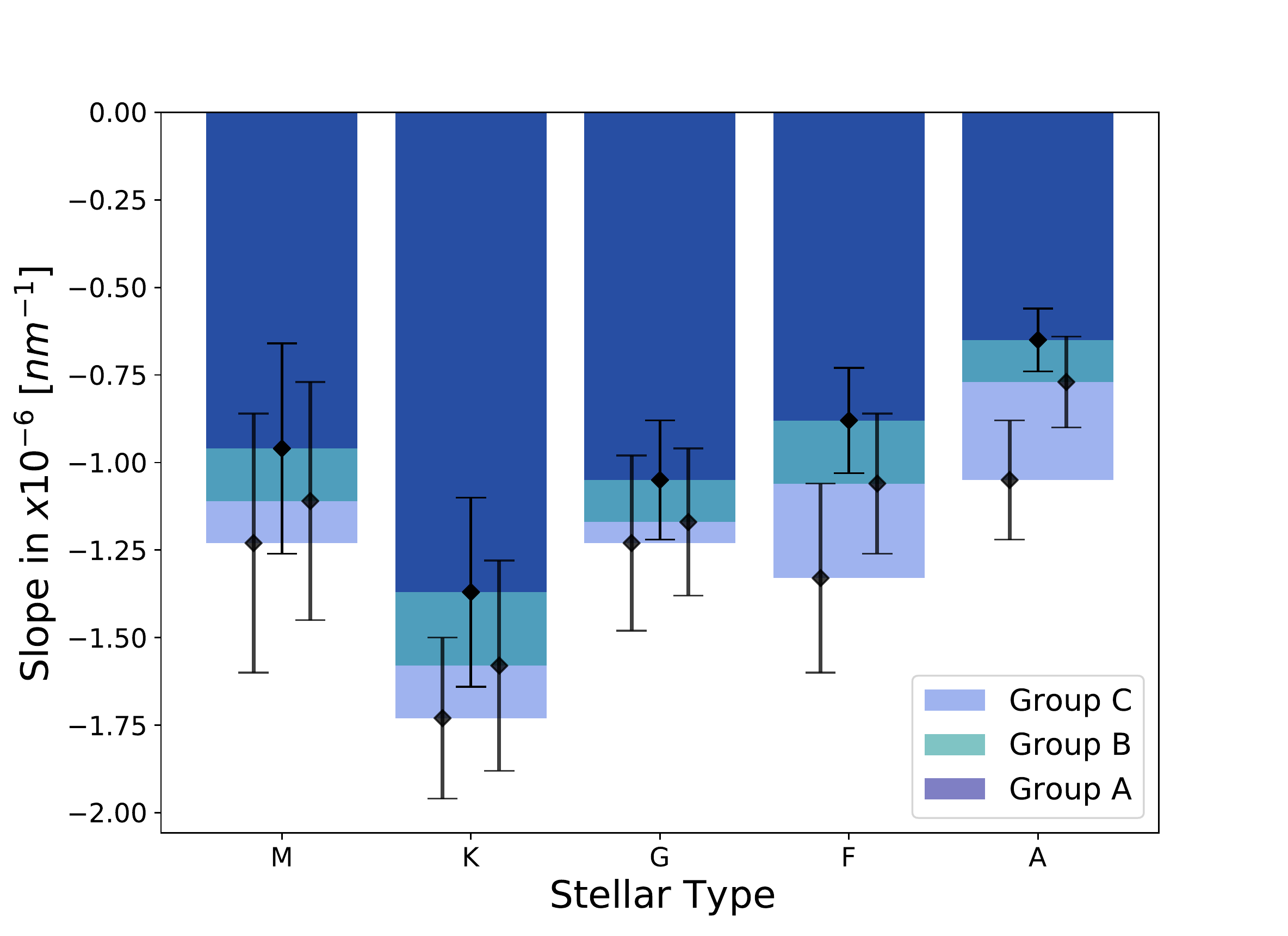}
\caption{Comparison of the different slope values between the different spectral types of host stars for $b+\Delta b$ of each subgroup. K-type stars exhibit slightly steeper spectral slopes for the different subgroups of exoplanets. We offset the values of the x-axis arbitrarily for clarity.}
\label{slopes}
\end{figure}

\subsection{Application on different stellar types}

We focus on determining how the different host star properties, in terms of center-to-limb variations, change the derived transmission spectra. We used synthetic light curves with the same outline of orbital parameters as in Section 3.1 in order to investigate this scenario. We again considered a hot-Jupiter exoplanet with the same characteristics as previously. We investigated four additional categories to the former K-type host case: M-, G-, F-, and A- spectral types; this is a total of five different host categories. We addressed the LD effect using the four-parameter LDL and coefficients from \cite{Claret2011}, derived for a logg = 4.5, solar metallicity, and for approximately a mean value of the different effective temperatures of each category of host stars ($T\textsubscript{eff}_M$=3800 K, $T\textsubscript{eff}_G$ = 5600 K, $T\textsubscript{eff}_F$ = 6250 K, and $T\textsubscript{eff}_A$ = 7500 K). All the parameters were kept fixed, except  for$R_p/R_s$.  
We investigated the formation of the error envelope of the same subgroups as in Section 3.2 by adopting a different host star. The resulting transmission spectra of the different subgroups are shown in Fig. \ref{diff_spectral_type_hosts}. The effect for A-type stars is slightly weaker than in M-type hosts. Interestingly, for M-type stars, we observe a feature at the blue wavelengths at 500nm that is persistent in all subgroups and might be linked to the wavelength dependence of the stellar LD effect. This feature is less pronounced in G-type hosts and A-type stars. F-type host stars exhibit the same shape as K-type stars. The derived slopes of each category (black dashed lines) are relatively similar, as is shown for $b+\Delta b$ in Fig.\ref{slopes}, for example. However, for K-type host stars, the spectral slopes are steeper than in the other stars. In addition, a linear trend progresses from K-type host stars that is interrupted at G-type hosts stars and continues to F- and A- type hosts for all groups (A,B, and C) of transiting exoplanets.

\section{Discussion}
Some discrepancies have been reported concerning the slope at optical wavelengths in the atmospheric characterization of exoplanets. An explanation for these inconsistencies can be the impact parameter degeneracy with spectral slope, for instance, the case of HAT-P-12b in \cite{Alexoudi2018}. Our work is able to show whether this controversy can be solved by a homogeneous set of orbital parameters. 

\begin{figure}[ht]
\centering
\includegraphics[height=9cm]{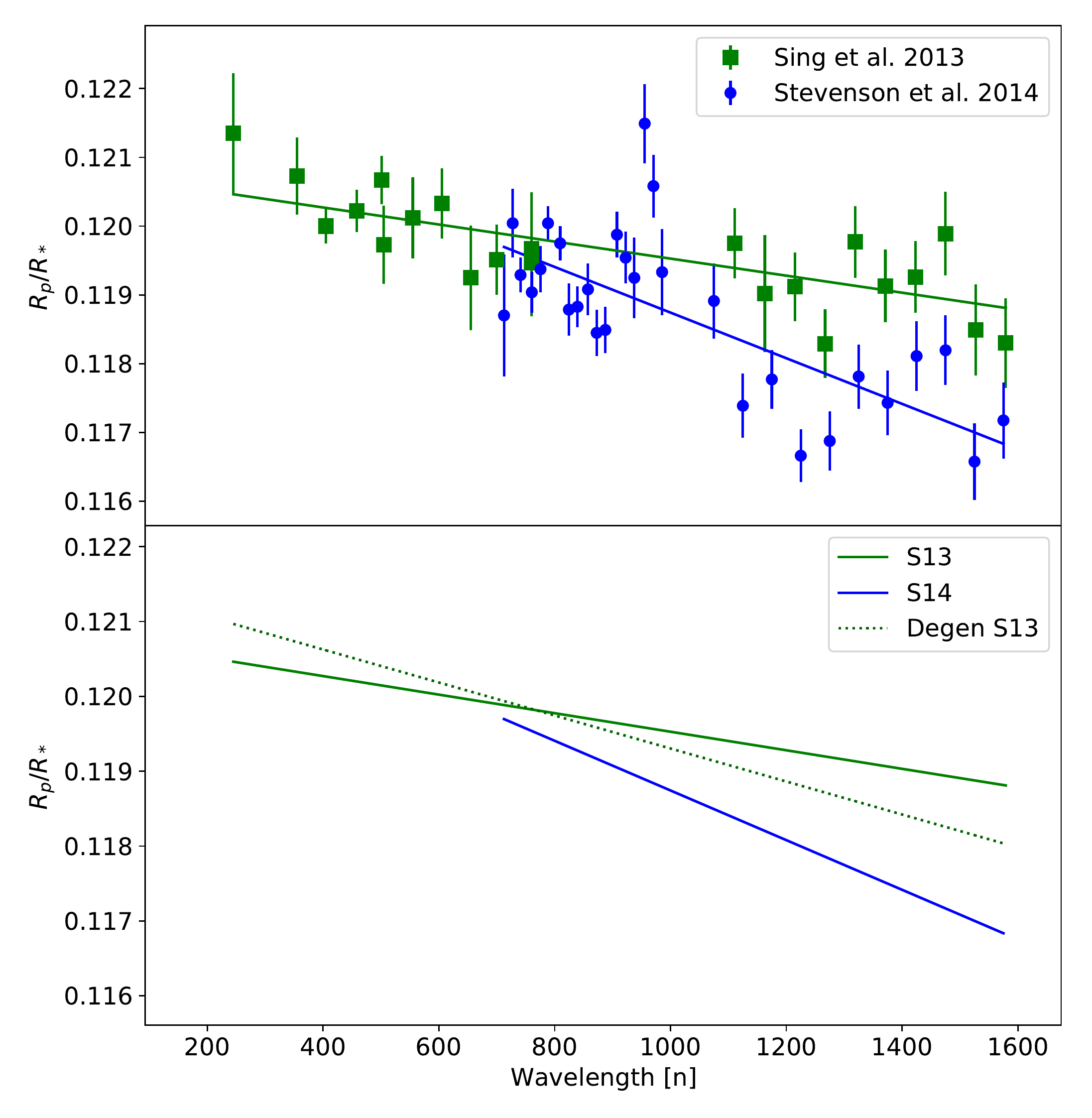}
\caption{Discrepant case of WASP-12b investigations (upper panel) in the works of S13 (green squares) and S14 (blue dots). We fit the original spectra of S13 and S14 with regression line fits as solid green and blue lines, respectively.  With our simulations (lower panel), we demonstrate the impact parameter degeneracy on the slope of S13 (green solid line). The synthetic light curves with the parameterization of S13 are fit with orbital parameters that yield the impact parameter of the S14 slope (solid blue line). Under this hypothesis, the S13 slope would be modified by an amount equal to the degenerated S13 slope (Degen S13, dotted green line).}
\label{wasp12b_comparison}
\end{figure}

\subsection{Case of WASP-12b}
We examined whether the discrepancy regarding two investigations on the atmospheric characterization of the ultra-hot Jupiter WASP-12b can be explained with the impact parameter degeneracy. \cite{Sing13} (hereafter S13) demonstrated using the Space Telescope Image Spectrograph (STIS) on the Hubble Space Telescope (HST), that the transmission spectrum of this exoplanet shows a Rayleigh signature at the blue wavelengths. However, \cite{Stevenson14} (hereafter S14), using ground-based data from the Gemini Multi-Object Spectrograph (GMOS), determined a different impact parameter for the system and concluded that this is a spectrum with a much steeper slope than did S13 (Fig. \ref{wasp12b_comparison}).  
We investigated the impact parameter degeneracy by creating synthetic UBVRIJH light curves with the lower impact parameter value $b=0.39$ from S13 and model-fit them with the fixed value of $b=0.48$ from S14. We treated the LD effect using the four-parameter LDL and coefficients from \cite{Claret2011}, using the ATLAS model and the stellar characteristics for WASP-12 of S13 ($T_{eff}$ = 6500, log$g$ = 4.5, [Fe/H] = 0.0). As expected from our results in the previous sections, the higher $b$ value of S14 results in a negative slope that is due to the impact parameter degeneracy, which we approximate by a linear regression line. We provide the linear slope $y$ of the published transmission spectra of S13 and S14 in the third column of Table.~\ref{tab_Fig2}, while the slope $y'$ that is caused by the two different impact parameter values is given in the fourth column. The lower panel of Fig. \ref{wasp12b_comparison} shows the two published slopes (solid lines) and the slope of S13 corrected for $y'$ (dashed line). The correction brings the slopes of the two WASP-12b transmission spectra into better agreement. We therefore conclude that the impact parameter degeneracy might contribute significantly to the difference in the published planetary spectra.

\begin{figure}[ht]
\centering
\includegraphics[height=8cm]{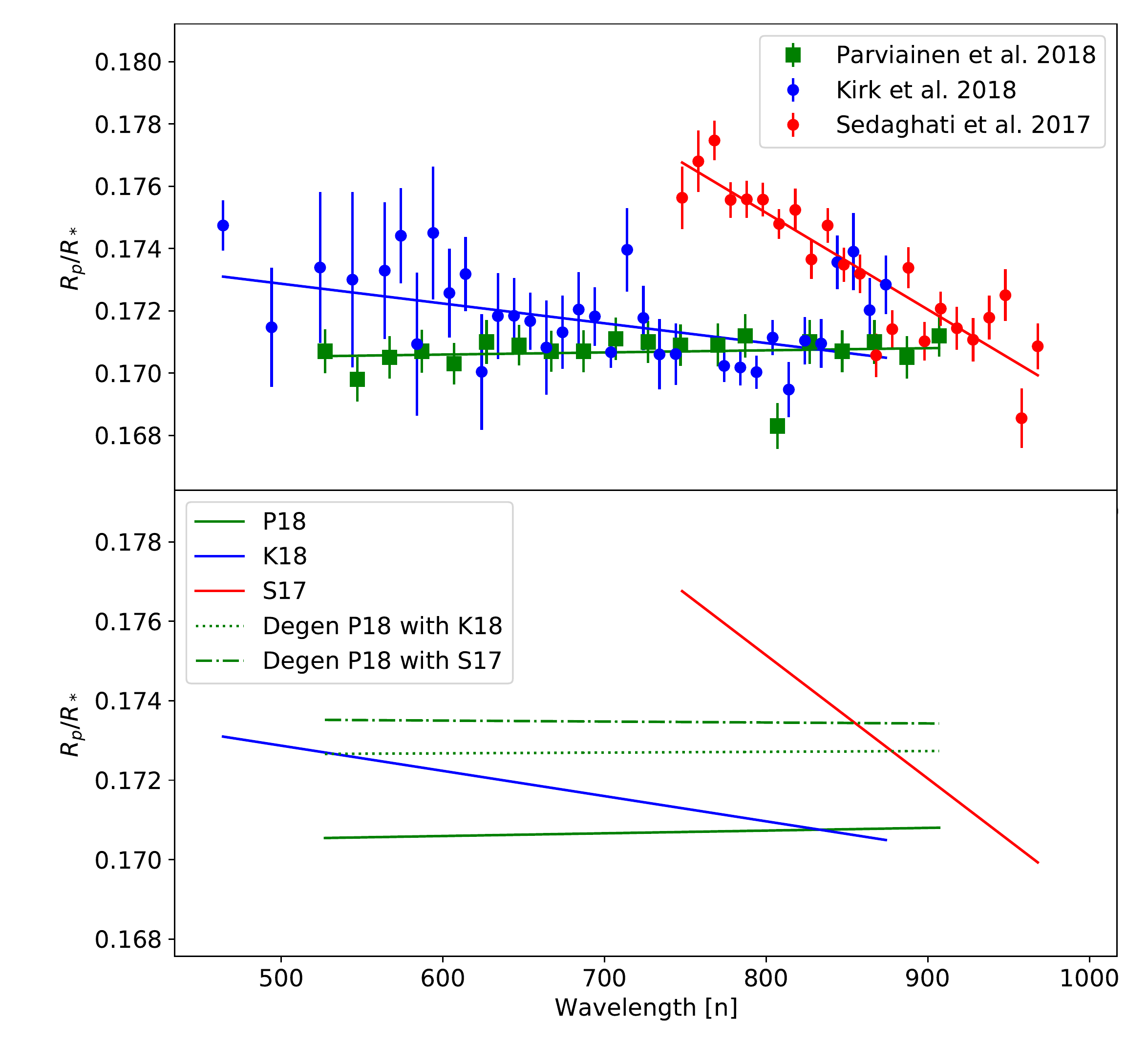}
\caption{
Spectral slopes from transmission spectroscopy investigations on WASP-80b by three individual groups (upper panel; P18 as green squares, K18 as blue dots, and S17 as red dots). The corresponding solid lines are regression line fits to each dataset. Synthetic light curves fit (lower panel) based on the parameters of P18 with transit models using the parameterization from K18 and S17, which yield different impact parameters than P18. The dotted green and dash-dotted green lines show the impact parameter degeneracy applied to the slope of P18, using the $b$ terms from K18 and S17, respectively. }
\label{wasp80b_comparison}
\end{figure}

\subsection{Controversial results for WASP-80b}
\cite{sedaghati2017} (hereafter S17), using data from the FOcal Reducer and low dispersion Spectrograph (FORS) on the Very Large Telescope (VLT), reported a ground-based transmission spectrum of WASP-80b showing a pronounced optical slope, which the authors interpreted as a spectral signature of potassium (K). For the same exoplanet, \cite{kirk2018} (hereafter K18), using the Auxiliary-port CAMera (ACAM) on the William Herschel Telescope (WHT), concluded that it has an atmosphere that is dominated by haze based on a mild slope in their transmission spectrum, and they reported a non-detection of the previous potassium claim. A third work by \cite{parviainen2018} (hereafter P18), using the Optical System for Imaging and low-Intermediate-Resolution Integrated Spectroscopy (OSIRIS) on the Gran Telescopio Canarias (GTC), yielded a flat spectrum that is indicative of high-altitude clouds for the atmosphere of WASP-80b. All three investigations fixed the impact parameter in their transit light-curve fit to different values. We study here whether these different assumptions can explain the different optical spectral slopes as a result of the impact parameter degeneracy. We created a set of UBVRIJH transit light curves with $b=0.16$, which is the value used by P18. We used the four-parameter LDL and adopted coefficients obtained from \cite{Claret2011}. Then we fit these synthetic light curves with $b$ fixed to 0.20 (K18), and in a second run to 0.23 (S17). The higher $b$ values of K18 and S17 compared to P18 cause a negative slope $y'$ by the impact parameter degeneracy, presented in Table.~\ref{tab_Fig2}. However, this value amounts to only a small fraction of the differences in the published slope values. 
A portion of the Rayleigh slope reported by K18 might therefore be attributed to the impact parameter degeneracy. The very different measured slope of S17 compared to K18 and P18 cannot be explained by the impact parameter degeneracy because the slope value caused by the deviating $b$ values in the synthetic spectra is negligible compared to the measured slope difference (Table~\ref{tab_Fig2}). Effects different from the degeneracy studied here therefore apparently dominate in this case.

\begin{figure}[ht]
\centering
\includegraphics[height=9cm]{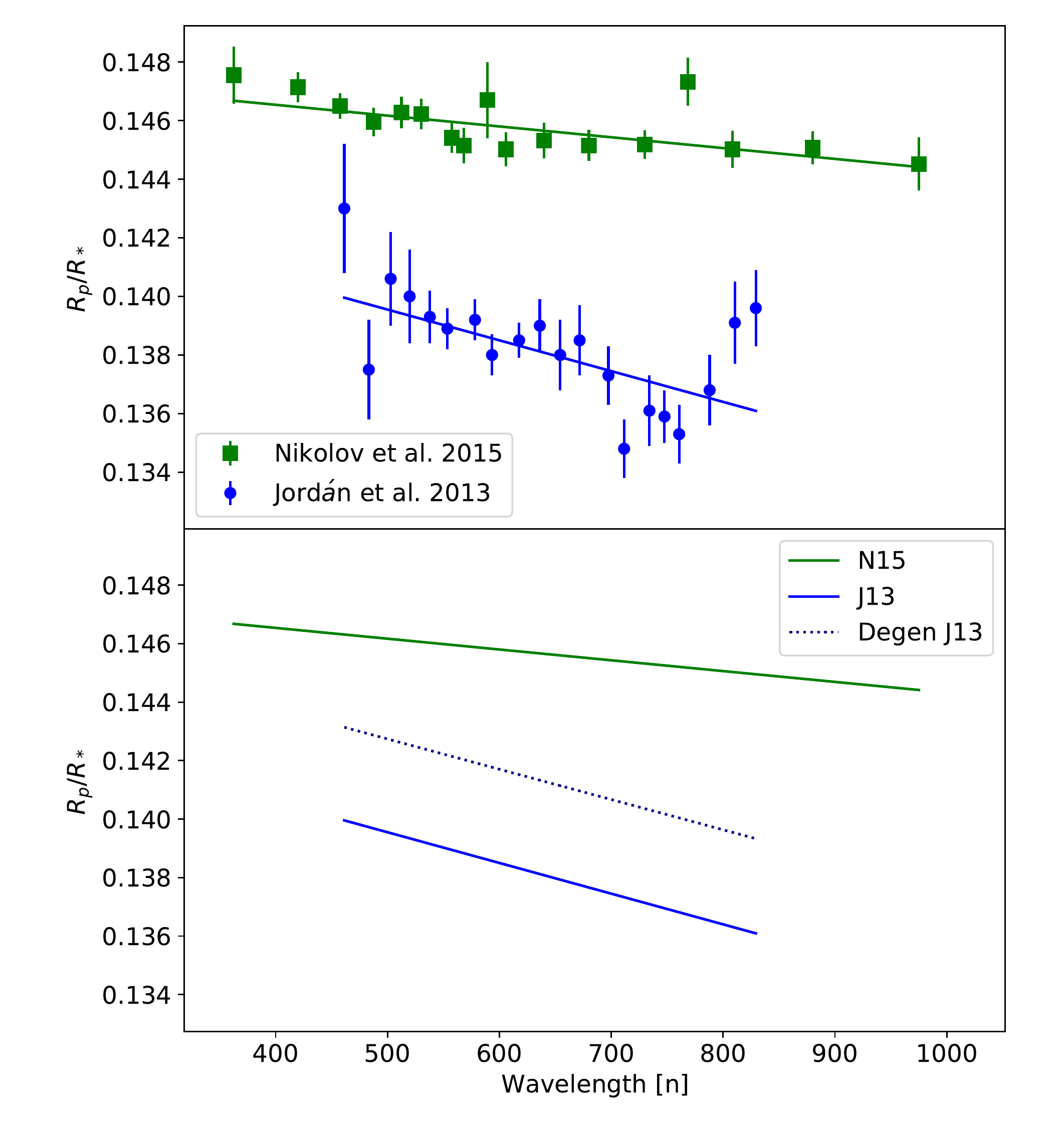}
\caption{WASP-6b transmission spectra (upper panel) of N15 and J13 as green squares and blue dots, respectively, along with the solid green line and solid blue line as regression line fits to these datasets. Synthetic light curves of J13 (lower panel) with parameters that yield the impact parameter of N15.  The contribution of this degeneracy to J13 is shown with a dotted blue line. }
\label{wasp6b_comparison}
\end{figure}

\begin{figure}[ht]
\includegraphics[height=9cm]{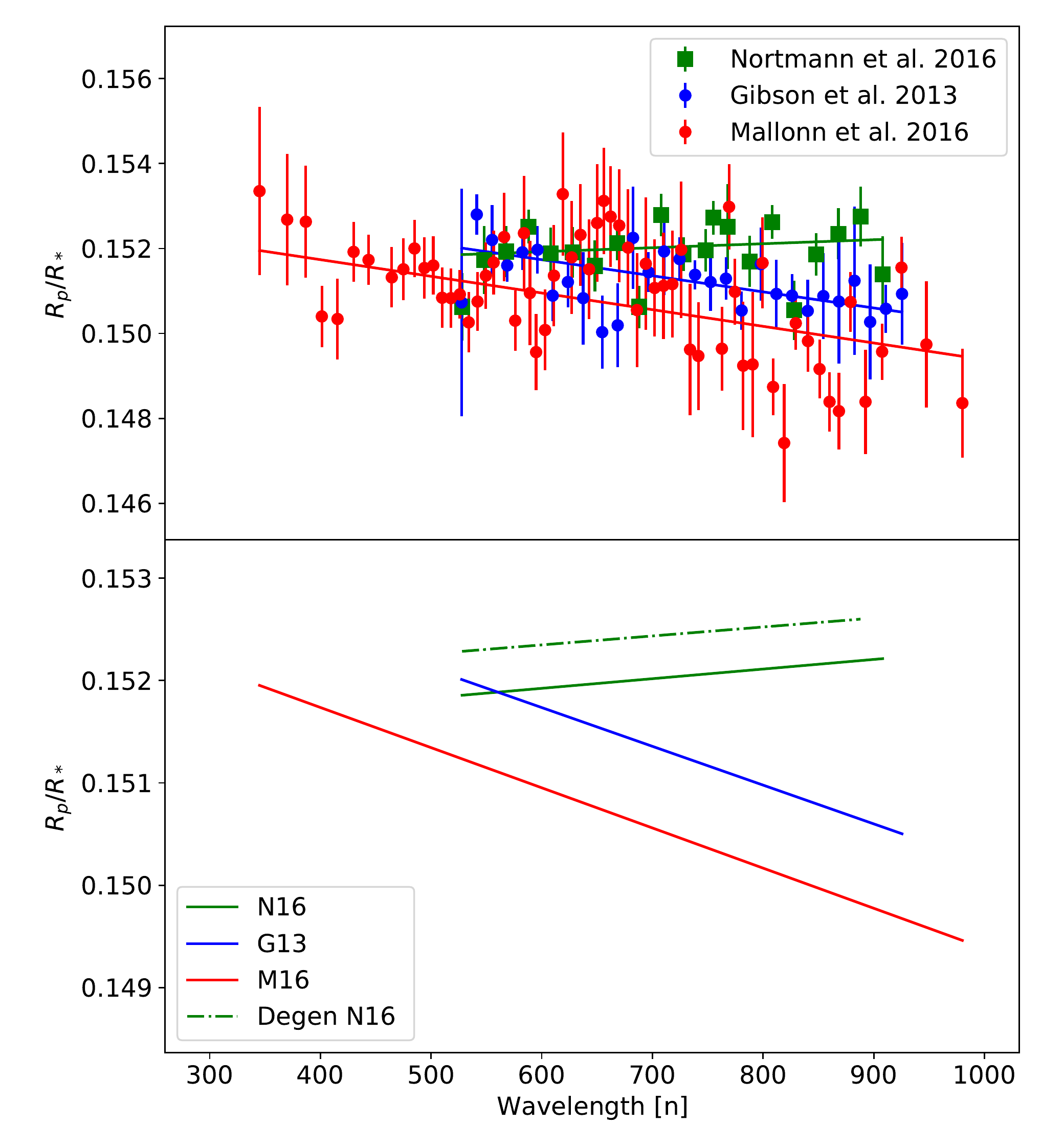}
\caption{HAT-P-32b transmission spectra (upper panel). We present results from N16 (green squares), G13 (blue dots) and M16 (red dots). The corresponding solid lines represent linear regression fits of the different works. Synthetic transmission spectrum of N13 (lower panel) using the impact parameter obtained from G13 and M16. The N16 slope is corrected for by this amount caused by the impact parameter degeneracy (dash-dotted green line).}
\label{hatp32b_comparison}
\end{figure}

\begin{table*}
\caption{Impact parameters from each study along with the gradient of the slope fit to the literature data ($y$) and the simulation slopes resulting from the impact parameter degeneracy ($y'$). By "input" we denote to the impact parameter value used to create synthetic light curves of a flat spectrum, and $y'$ is the slope caused by fitting these synthetic light curves with the deviating impact parameter.}
\label{tab_Fig2}
\begin{center}
\begin{tabular}{ccccc}

\hline
 Planet &Reference &$y\,[\times 10^{-6}$~nm$^{-1}]$ & $y'\,[\times 10^{-6}$~nm$^{-1}]$ &$b$ \\
\hline
WASP-12b  &\cite{Sing13} &-1.24$\,\pm\,$0.25 &input &0.39 \\
         &  \cite{Stevenson14} &-3.32$\,\pm\,$0.62 &-0.97  &0.48 \\
    \hline    
WASP-80b  &  \cite{parviainen2018}&0.68$\,\pm\,$1.28 &input&0.16 \\
&  \cite{kirk2018} &-6.35$\,\pm\,$2.08 &$-0.49$   &0.20 \\
        &   \cite{sedaghati2017} &-31.05$\,\pm\,$3.30 &$-0.92$  &0.23 \\
    \hline
HAT-P-32b & \cite{Nortmann16} &0.94$\,\pm\,$1.32  &input &0.07 \\
          & \cite{Gibson13}   &-3.79$\,\pm\,$0.94 &-0.07   &0.09 \\
          & \cite{MallonnStrass}&-3.92$\,\pm\,$0.94 &-0.07   &0.09 \\
          \hline
WASP-6b & \cite{Jordan2013}   &-10.56$\,\pm\,$3.28  &input &0.26 \\
          & \cite{Nikolov2015}  &-3.69$\,\pm\,$1.11 &-0.14   &0.28\\
     
\hline 
\end{tabular}
\end{center}
\end{table*}

\subsection{Inexplicable cases}
The impact parameter degeneracy can be an explanation for parts of the reported discrepancies regarding the atmospheric characterization of exoplanets with respect to their spectral slopes. However, it is fails to clarify the inconsistencies in centrally transiting exoplanetary systems; especially when the individual analyses make use of a quite similar $b$ value. For instance, the discrepancy on WASP-6b between \cite{Jordan2013} and \cite{Nikolov2015} (Fig.~\ref{wasp6b_comparison}) cannot be explained. The two groups used similar $b$ values, and their small difference causes only a very small slope $y'$ when tested with synthetic light curves (Table~\ref{tab_Fig2}). Therefore the different amplitudes of the discovered Rayleigh feature cannot be attributed to the impact parameter degeneracy.

Another case is HAT-P-32b. This exoplanet has been studied thoroughly in the literature \citep[e.g.,][]{Gibson2013,MallonnStrass,Nortmann16,TregloanReed2018,Alam2020}. However, some investigations exhibit significant differences in the spectral slope of the obtained transmission spectra. We compare the results of \cite{Gibson13} (hereafter G13) and \cite{MallonnStrass} (hereafter (M16) with the result of \cite{Nortmann16} (hereafter N16). The first two studies achieve transmission spectra with a negative low-amplitude slope that might indicate scattering processes in the planetary atmosphere, and the result of N16 supports the scenario of a very flat spectrum. M16 used the same orbital parameter values as G13 in their light-curve analysis, while N16 used values resulting in a slightly different $b$ value (Fig. \ref {hatp32b_comparison}). When we created synthetic light curves with the $b$ value of N16 and fit them with $b$ fixed to the value of G13 and M16, we obtained a slope caused by the impact parameter degeneracy of negligible gradient (Table~\ref{tab_Fig2}). This follows our findings in previous sections that the impact parameter degeneracy is less important for centrally transiting systems and for a precisely determined impact factor $b$.
Thus, the impact parameter degeneracy is certainly not the source of the deviating results of N16 and G13, and M16, on the optical spectral slope.

\subsection{Impact parameter degeneracy versus other causes of spectral slope uncertainty}

The case-by-case investigations of individual systems presented in the former sections showed that the impact parameter degeneracy is certainly not the only effect that can generate uncertainties in the optical slope of exoplanet transmission spectra. Many studies on the effect of dark or bright spots in the stellar photosphere have been conducted, which in the case of very active stars can cause optical slopes of larger amplitudes than the impact parameter degeneracy \citep{McCullough2014,oshagh2014,rackham2018,Mallonn2018}. However, for stars at about the low activity level of the Sun, the effect might be negligible. Third-light contribution of another star in the photometric aperture can also mimic a spectral slope. If it is uncorrected for, the amplitude of this effect can be stronger than the impact parameter degeneracy \citep[e.g.,][]{Sing13,MallonnStrass,vonEssen2020}. However, the Gaia satellite astrometry mission \citep{GaiaDR2} has provided information on significant foreground or background objects to the exoplanet host stars, and a third-light correction can be performed with good accuracy. The choice of the stellar LD law or the estimation of the LD coefficients might also affect the planetary spectral slope. We tested the amplitude of this effect by creating synthetic UBVRIJH light curves with the four-parameter LD law and fit these noise-free data with transit models using the two-parameter quadratic LD law or the one-parameter linear law. We did not apply deviations from the orbital parameters, but kept them fixed to their input values. The only parameters left free to vary were the ratio of the planet-to-star radius. We derived values that deviated from the input value by an order of magnitude less than the impact parameter degeneracy for a typical uncertainty on $b$ (Section~\ref{sec_res2}).

Several of the individual planets we described above with discrepant published spectral slopes are inactive and either do not have a known third-light contribution, or a third-light correction has been performed using similar correction values. This means that none of the listed effects can explain the slope discrepancies. Our list of potential sources therefore appears to be incomplete, and other reasons such as different light-curve detrending approaches or systematics in the observing data might also play a role. For example, \cite{Nikolov2015} and \cite{Stevenson14} compared their result on WASP-6b, respectively WASP-12b, to the previously published results of \cite{Jordan2013} and \cite{Sing13}, respectively, and argued that differences in the employed systematics model are a likely reason for an offset in $R_p/R_s$. We speculate that an offset like this might not be entirely achromatic, but wavelength dependent, and also affect the measured slope. The controversial results for WASP-6b, HAT-P-32b, and WASP-80b, for example,  indicate that these systematics-related effects on the measured slope can be stronger than the impact parameter degeneracy.

\section{Summary and conclusions}

The limited precision in the determination of $b$ in turn limits the characterization of exoplanetary atmospheres. We addressed the degeneracy of the spectral slope with this parameter through two main investigations, using synthetic noise-free light curves. First with a qualitative approach, in order to demonstrate that the changes in $\Delta b$ affect the direction of the spectral slope, and then with a quantitative investigation to determine the error envelope of this effect for different groups of exoplanets, for which we applied typical measurement uncertainties in $b$. We conclude that the impact parameter degeneracy can be the driver of the spectral slope in both directions (positive and negative slopes), and it can transform flat spectra into sloped spectra, and vice versa. The effect persists with the use of different stellar hosts and yields steeper slopes for K-type hosts, but introduces a feature at the bluer wavelengths for M-type hosts. 

The amplitude of the slope caused by the impact parameter degeneracy for a typical uncertainty in $b$ is about one scale height over the optical wavelength range for a representative inflated hot Jupiter with a comparably large scale height suitable for transmission spectroscopy. For planets with smaller scale heights and therefore potentially weaker transmission spectroscopy signals, the amplitude of the impact parameter degeneracy amounts to even higher values in units of the scale height. Typical reported spectral slopes measured from observations are one to three scale heights in amplitude, therefore we consider the impact parameter degeneracy to be able to affect the measurements significantly.

We discussed the application of the degeneracy on a sample of reported discrepancies from the literature, but found no planet next to HAT-P-12b \citep{Alexoudi2018} for which the impact parameter degeneracy can fully explain the differences between reported optical slopes of its transmission spectrum. For WASP-12b, the degeneracy might partly be responsible for a reported discrepancy, but there are several other systems, for instance, WASP-80b, WASP-6b, or HAT-P-32b, for which the amplitude of a potential impact parameter degeneracy is negligible compared to the amplitude of the reported discrepancy. This illustrates that there is more than one source of error for the optical slope in exoplanet transmission spectroscopy.  

As a consequence of the results, we suggest that the orbital parameter is not kept fixed in a model fit of the chromatic light curves when transmission spectra are extracted. We instead advise to let it remain a free parameter, potentially constrained by Gaussian or uniform priors. Another possibility is performing a similar exercise as done in this work and fixing the impact factor in a first run to its best-fit value, and compare the outcome of the transmission spectrum in a second run when the impact factor is changed by an uncertainty of about one sigma.

\begin{acknowledgements}
We sincerely thank the anonymous referee for the valuable remarks and comments, which significantly contributed to the quality of our paper.
This research has made use of the NASA Exoplanet Archive, which is operated by the California Institute of Technology, under contract with the National Aeronautics and Space Administration under the Exoplanet Exploration Program.
XA is grateful for the financial support from the Potsdam Graduate School (PoGS) in form of a doctoral scholarship. 
\end{acknowledgements}

\bibliographystyle{aa}
\bibliography{main}
\end{document}